\documentclass[aps,pre,twocolumn,superscriptaddress,showpacs,preprintnumbers,amsmath,amssymb]{revtex4}
\usepackage{graphicx}
\graphicspath{{fig/}}
\usepackage{dcolumn}
\usepackage{bm}
\usepackage{graphicx}
\usepackage{amssymb}
\usepackage{amsmath}
\usepackage{color}

\def\bea{\begin{eqnarray}}
\def\eea{\end{eqnarray}}
\def\ba{\begin{eqnarray}}
\def\ea{\end{eqnarray}}
\def\beq{\begin{eqnarray}}
\def\eeq{\end{eqnarray}}
\def\be{\begin{equation}}
\def\ee{\end{equation}}

\def\bm{\begin{math}}
\def\me{\end{math}}

\begin{document}

\preprint{APS/123-QED}

\title{Velocity distribution function  and effective constant restitution coefficient
for granular gas of viscoelastic particles}
\author{Awadhesh Kumar Dubey}%
\affiliation{%
School of Physical Sciences, Jawaharlal Nehru University, New Delhi, 110067, India.
}%
\author{Anna Bodrova}%
\affiliation{%
Faculty of Physics, M.V.Lomonosov Moscow State University, Moscow, 119991, Russia
}%

\author{Sanjay Puri}%
\affiliation{%
School of Physical Sciences, Jawaharlal Nehru University, New Delhi, 110067, India.
}%
\author{Nikolai Brilliantov}%
\affiliation{%
Department of Mathematics, University of Leicester, Leicester LE1 7RH, United Kingdom
}%

\date{\today}

\begin{abstract}

We perform large-scale event-driven Molecular dynamics (MD) simulations for granular
gases of particles interacting with the impact-velocity dependent restitution
coefficient $\varepsilon (v_{\rm imp})$. We use $\varepsilon (v_{\rm imp})$ as it
follows from the simplest first-principle collision model of viscoelastic spheres. Both
cases of force-free and uniformly heated gases are studied. We formulate a simplified
model of an effective constant restitution coefficient $\varepsilon_{\rm eff}$, which
depends on a current granular temperature and compute $\varepsilon_{\rm eff}$, using the
kinetic theory. We develop a theory of  the velocity distribution function for driven
gases of viscoelastic particles and analyze evolution of granular temperature and of
the Sonine coefficients, which characterize the form of the velocity distribution
function.  We observe that for not large dissipation the simulation results are in an
excellent agreement with the theory for both, homogeneous cooling state and uniformly
heated gases.  At the same time a noticeable discrepancy between the theory and MD
results for the Sonine coefficients is detected for large dissipation. We analyze the
accuracy of the simplified model, based on the effective restitution coefficient
$\varepsilon_{\rm eff}$ and conclude that this model can accurately describe granular
temperature. It provides also  an acceptable accuracy for the velocity distribution
function for small dissipation, but fails when dissipation is large.
\end{abstract}

\pacs{81.05.Rm, 05.20.Dd, 05.40.2a}

\maketitle


\section{Introduction}
\label{intro} Granular materials are abundant in nature; possible examples range from
sands and powders on Earth \cite{Herr1998,jnb96,hw04,r00,d00} to more dilute systems,
termed as granular gases \cite{PL2001,PB2003,bp04,g03},  in space. Astrophysical
objects, like planetary rings or dust clouds are typical examples of granular gases
\cite{PL2001,Brahic1984}. Depending on the average kinetic energy of individual grains,
granular systems can be in a solid, liquid or gaseous state. In the present study we
address a low-density granular gas -- a system comprised of macroscopically large
number of macroscopic grains that  move ballistically between inelastic pair-wise
collisions. The inelastic nature of inter-particle collisions is the main feature which
distinguishes a granular gas from a molecular gas and gives rise to many astonishing
properties of dissipative gases. If no external forces act on the system it evolves
freely and permanently cools down. During the first stage of a granular gas evolution
its density remains uniform and no macroscopic fluxes present. This state, which  is
called a {\it homogeneous cooling state} (HCS) \cite{bp04,h83}, is however, unstable
with respect to fluctuations and in later times clusters and vortices develop. The
system then continue to loose its energy in the {\it inhomogeneous cooling state} (ICS)
\cite{gz93,dp03,dppre03,n03,nebo79,neb98,ap06,ap07}.

If energy is injected into a granular gas to compensate its losses in dissipative
collisions, the system settles into a nonequilibrium steady-state
\cite{Sant2000,vne98}. There are different ways to put energy into the system: via
shearing \cite{shear}, vibration \cite{wildman,menon} or rotating \cite{rotdriv,puha}
of the walls of a container, applying  electrostatic \cite{electrodriven} or magnetic
\cite{magndriven} forces, etc. In planetary rings the energy losses due to particles'
collisions are replenished by gravitational interactions \cite{spahn1997,ringbook}. To
describe theoretically the injection of energy  into a system a few types of
thermostat, which mimics an action of external forces, have been proposed
\cite{Sant2000}. Here we exploit  a white-noise thermostat \cite{wm}, where all
particles are heated uniformly and independently (see the next sections for more
detail).

Energy dissipation in a pair-wise collision is quantified by the restitution
coefficient $\varepsilon$,
\ba \label{rc} \varepsilon = \left|\frac{\left({\bf
v}^{\,\prime}_{12} \cdot {\bf e}\right)}{\left({\bf v}_{12} \cdot {\bf
e}\right)}\right| \, ,
\ea
where ${\bf v}^{\,\prime}_{12}={\bf v}_{2}^{\,\prime}-{\bf v}_{1}^{\,\prime}$ and ${\bf
v}_{12}={\bf v}_{2}-{\bf v}_{1}$ are the relative velocities of two particles after and
before a  collision, and ${\bf e}$ is a  unit vector connecting their centers at the
collision instant. The post-collision velocities are  related to the pre-collision
velocities ${\bf v}_{1}$ and ${\bf v}_{2}$ as follows, e.g. \cite{bp04}:
\begin{equation}
\label{v1v2} {\bf v}_{1/2}^{\,\prime} = {\bf v}_{1/2} \mp  \frac{1+\varepsilon}{2}({\bf
v}_{12} \cdot {\bf e}){\bf e} \, .
\end{equation}
To date, most studies of the HCS and ICS have focused on the case of a  constant
restitution coefficient \cite{h83,gz93,dp03,dppre03,n03,nebo79,neb98,ap06,ap07}. This
assumption contradicts, however, experimental observations \cite{w60,bhd84,kk87}, along
with basic mechanical laws \cite{titt91,rpbs99}, which indicate  that $\varepsilon$
does depend on the impact velocity \cite{kk87,rpbs99,bshp96,mo97,sp98}. This dependence
may be obtained by solving the equations of motion for colliding particles with the
explicit account for the dissipative forces acting between the grains. The simplest
first-principle model of inelastic collisions assumes viscoelastic properties of
particles' material, which results in viscoelastic inter-particle force \cite{bshp96}
and finally in the restitution coefficient \cite{sp98,rpbs99}:
\begin{equation}
\varepsilon=1-C_{1}A\kappa^{2/5}\left|{\bf v}_{12} \cdot {\bf e}\right|^{1/5}
+C_{2}A^{2}\kappa^{4/5}\left|{\bf v}_{12} \cdot {\bf e}\right|^{2/5} \ + ... \ .
\label{epsx}
\end{equation}
Here, the numerical coefficients are $C_{1}\simeq 1.15$ and  $C_{2} \simeq 0.798$.  The
elastic constant \ba \label{kappa} \kappa = \left (\frac{3}{2}\right
)^{3/2}\frac{Y\sqrt{\sigma}}{m(1-{\nu}^2)} \ea
is a function of the Young's modulus $Y$, Poisson ratio $\nu$,  mass $m$ and diameter
$\sigma$ of the particles; the constant $A$ quantifies the viscous properties of the
particles' material. While the last equation is valid for not very large inelasticity,
in the recent work of Schwager and Poschel \cite{sp08} an expression for $\varepsilon$,
which is valid for high inelasticity and accounts for the delayed recovery of the shape
of colliding particles has been derived:
\ba \label{rerc} \varepsilon = 1 + {\sum^{N_{\varepsilon}}_{k=1}}   h_k{\delta}^{k/2}
\left[2u\left ( t\right )\right ]^{k/20}\left |\left ({{\bf c}}_{12}\cdot{\bf e}\right
)\right |^{k/10} \, , \ea
where $h_{1}=0$, $h_{2}=-\, C_{1}$, $h_{3}=0$,  $h_{4}=C_{2}$, and other numerical
coefficients up to $N_{\varepsilon}=20$ are given in Ref. \cite{sp08}. In the latter
equation $u(t) = T(t)/T_0$ ($T_0 = T(0)$), with the usual definition of temperature
$T(t)$,
\begin{equation}
\frac{3}{2}nT(t)=\int d{\bf v}~\frac{m{\bf v}^{\, 2}}{2}f({\bf v},t) \, . \label{grantemp}
\end{equation}
Here  $n$ is the number density of the gas, ${\bf v}$ is a particle velocity, $f({\bf
v},t)$ --  the time-dependent velocity distribution function and ${\bf c}_{12}$ is the
dimensionless relative velocity, defined as ${\bf c}_{12}= {\bf v}_{12}/{v_T}$,  where
$v_T(t) = \sqrt{2T(t)/m}$ is the thermal velocity. The dissipation constant $\delta$ in
Eq.~(\ref{rerc}) reads,
\ba \label{delta} \delta = A\kappa^{2/5}
\left(\frac{T_0}{m}\right)^{1/10}, \ea
so that the first few terms in Eq.~(\ref{rerc}) are identical to those in
Eq.~(\ref{epsx}).  Note that $\delta = 0$ corresponds to an elastic system,
$\varepsilon =1$ and dissipation in the system increases with increasing $\delta$.
Furthermore, the restitution coefficient for all collisions increases and tends to
unity, $\varepsilon \rightarrow 1$, when the (reduced) temperature decreases, that is,
when a granular gas cools down.

Note that although we address here dry granular particles, they  still can stick in collisions with a very
small impact velocity. Namely, if the normal component of the relative velocity $\left| {\bf v}_{12}\cdot
{\bf e} \right|$ is smaller than the sticking threshold,
$$
g_{\rm st} = \sqrt{\frac{4q_0 \left(\pi^5 \gamma^5 D^2 \right)^{1/3}}{m \varepsilon^2}
\left(\frac{\sigma }{4} \right)^{4/3}} \,,
$$
where $D = 3(1-\nu^2)/2Y$, $q_0 \simeq 1.45$ and $\gamma$ is the adhesion coefficient, the colliding
particles cling together \cite{adh2007, adh2006}; in this case the restitution coefficient drops to zero. In
the present study we neglect these processes.

The velocity distribution function in granular gases  usually deviates from a
Maxwellian distribution \cite{ap06,ap07,gs95,ep97,vne98} and may be described by the
{\it Sonine-polynomial} expansion \cite{bp04}:
\begin{eqnarray}\nonumber
&&f({\bf v},t)= \frac{n}{v_{T}^{3}}\tilde{f}({\bf c},t) ,  \\
&&\tilde{f}({\bf c},t)\!=\!\pi^{-3/2} \exp \left(-c^{2} \right)
\left[1\!+\!\sum_{p=1}^{\infty}a_{p}(t)S_{p}(c^2)\right] \label{Sdistrib}
\end{eqnarray}
where  ${\bf c}={\bf v}/v_T$ and the Sonine polynomials read for $d=3$:
\begin{equation}
S_{p}(c^2) = \sum_{k = 0}^{p} \frac{(-1)^k(1/2+p)!}{(1/2+k)!(p-k)!k!} c^{2k} \, .
\end{equation}
Hence  evolution of $f({\bf v},t)$ is completely determined by the time dependence of the Sonine coefficients
$a_p(t)$. The first Sonine coefficient is trivial, $a_1=0$, as it follows from the definition of temperature,
e.g. \cite{bp04}; in the elastic limit, that is, for a  Maxwellian distribution, all Sonine coefficients are
zero, $a_p=0$.
The Sonine coefficients characterize the successive moments $\langle
c^{2k}\rangle$ of the velocity distribution function (\ref{Sdistrib}), so that the first few moments read,
\begin{eqnarray}
\label{cak1}
\left\langle c^2 \right\rangle &=& \frac{3}{2} , \\
\qquad \left\langle c^4 \right\rangle &=& \frac{15}{4}\left(1+a_2\right) , \nonumber\\
\qquad \left\langle c^6 \right\rangle &=& \frac{105}{8}\left(1+3a_2-a_3\right)
.\nonumber
\end{eqnarray}

It has been shown recently that for a granular gas with a constant $\varepsilon$ the
expansion (\ref{Sdistrib}) converges for small and moderate dissipation, up to
$\varepsilon \gtrsim 0.6$, but breaks down otherwise \cite{bp06}. For a moderate
dissipation it is sufficient to consider only the  first two non-trivial coefficients
$a_2$ and $a_3$, which have been studied both analytically
\cite{gs95,vne98,bp06,Sant2009} and by means of computer simulations
\cite{ap06,ap07,n03,bp06,hob00,bcr96,Sant2009}.

The theory developed for a granular gas with a constant $\varepsilon$ in a homogeneous
cooling state  predicts  a rapid relaxation of $a_p$ to a steady state values, which
depend on $\varepsilon$ \cite{vne98,bp06}. After a certain time,  spatial structures,
like  vortices and clusters, spontaneously arise in such a granular gas
\cite{gz93,dp03,dppre03,n03,nebo79,neb98,ap06,ap07}. The velocity distribution function
becomes effectively Gaussian, which happens due to an averaging over a large number of
independent clusters, where particles' velocities  are approximately parallel; $a_p$,
however, show very strong fluctuations around zero mean values \cite{ap06,ap07}. In a
granular gas of viscoelastic particles clusters and vortices appear only as transient
phenomena \cite{BrilPRL2004}; moreover cluster formation is completely suppressed in a
granular gas, orbiting a massive body in its gravitational potential -- a typical
example of such systems is a Kepler-disc in astrophysics \cite{spahn1997}.

In contrast to the case of a constant $\varepsilon$, evolution of the velocity
distribution function in a gas of viscoelastic particles demonstrates a complicated
non-monotonous time dependence \cite{bp00,bb09}. Typically, the magnitudes of the
Sonine coefficients first increase with time, reach maximal values, and finally tend to
zero, so that asymptotically the Maxwellian distribution is recovered \cite{bp00,bb09}.
This simply follows from the fact that the collisions become more and more elastic as
temperature decreases. Although the time-dependence of $a_2(t)$ and
$a_3(t)$ in a force-free granular gas of viscoelastic particles has been studied
theoretically in \cite{bb09}, the respective analysis for heated gases is still
lacking.  Moreover, evolution of the velocity distribution function and its properties
in such gases have been never studied numerically, by means of MD, neither for
force-free nor for driven systems. Hence, it is still not known, how accurate is the
theory for the velocity distribution function in this case and what is the range of
convergence of the Sonine series with respect to dissipation parameter $\delta$.

Another interesting question refers to  the possibility of a simplified description of
a granular gas of viscoelastic particles, using an effective constant restitution
coefficient: While  it is known that the model of a constant $\varepsilon$ fails
qualitatively to describe a complicated evolution of such gas in a homogeneous cooling
state \cite{BrilPRL2004}, it is still not clear, how accurate could be the respective
description for the case of a driven system. Also, it would be interesting to check,
whether a model of "quasi-constant" restitution coefficient, where $\varepsilon$ does
not depend on an impact velocity, but depends on current temperature, may be used for a force-free gas.

In the present study we perform thorough large-scale event-driven MD simulations of the
velocity distribution function in a gas of viscoelastic particles, both in a
homogeneous cooling state and under a uniform heating. We develop a theory for $f({\bf
v},t)$ for driven gases and observe that our MD results for $a_2(t)$ and $a_3(t)$ are
in an excellent agreement with theoretical predictions of \cite{bb09} for the HCS and
of the novel theory for driven systems. We report simulation results for the high-order
Sonine coefficients $a_4$ and $a_5$, which may be hardly obtained from the kinetic
theory. The analysis of these coefficients allows to asses the convergence of the
Sonine series. We also analyze the model of "quasi-constant" restitution coefficient
$\varepsilon_{\rm eff}$ and show that it can be used for an accurate description of a granular gas if dissipation is not large.

The paper is organized as follows. In Sec.~\ref{sec3}  we develop a theory for the velocity distribution
function for uniformly heated gases, that is,  we compute time dependent Sonine coefficients $a_2(t)$ and
$a_3(t)$. In Sec.~\ref{Sepsilon} we derive an effective "quasi-constat" restitution coefficient
$\varepsilon_{\rm eff}$. In Sec.~\ref{sec2} we report our MD results for a force-free and heated gas of
viscoelastic particles.  In Sec.~\ref{sec4} the theoretical predictions are compared with simulation results
and the accuracy of the theory, based on $\varepsilon_{\rm eff}$, is scrutinized. Finally, in Sec.~\ref{sec5}
we summarize our findings.

\section{Velocity distribution function for uniformly heated gas }
\label{sec3}

Assuming the molecular chaos, which is an adequate hypothesis for dilute granular gases addressed here, e.g.
\cite{bp04}, we can write the  Boltzmann-Enskog (BE) equation, supplemented by the diffusive Fokker-Planck
term, that mimics a uniform heating \cite{vne98}:
\be \frac{\partial f({\bf v},t)}{\partial t}=g_{2}(\sigma)I(f,f) \,
 + \frac{\xi^{2}_{0}}{2}\frac{\partial^2}{\partial\bf v^2}f({\bf v},t). \label{rbe}
\ee
Here $I(f,f)$ is the collision integral and $g_2(\sigma)$ is contact value of the pair
correlation function that takes into account the excluded volume effects \cite{bp04}.
For the scaling distribution function $\tilde{f}({\bf c},t)$ [cf. Eq.
(\ref{Sdistrib})], the above equation reads
\be \frac{\partial\tilde{f}}{\partial
t}-\frac{1}{v_{T}}\frac{dv_{T}}{dt}\left(3\tilde{f}+c\frac{\partial\tilde{f}}{\partial
c}\right)
=g_{2}\sigma^{2}nv_{T}\tilde{I}+\frac{1}{v^{2}_{T}}\frac{\xi^{2}_{0}}{2}\frac{\partial^{2}}{\partial{\bf
c}^{2}}\tilde{f},            \label{sbe} \ee
where $\tilde{I}\left(\tilde{f},\tilde{f}\right) = \sigma^{-2} n^{-2}
v_T^{2}I\left(f,f\right)$ is the dimensionless collision integral, defined as
\begin{eqnarray}
\tilde{I}(\tilde{f},\tilde{f})&=&\int d{\bf c}_{2} \int d{\bf e} \, \Theta (-{\bf c}_{12}\cdot {\bf e}\,
)
\left|-{\bf c}_{12} \cdot {\bf e}\, \right|  \label{II} \\
&\times& \left[\chi\tilde{f}({\bf c}_{1}^{\ \prime\prime},t)\tilde{f} ({\bf c}_{2}^{\
\prime\prime},t)-\tilde{f}({\bf c}_{1},t)\tilde{f}({\bf c}_{2},t)\right] \, , \nonumber
\end{eqnarray}
with ${\bf c}_{1}^{\, \prime\prime}$ and ${\bf c}_{2}^{\, \prime\prime}$ being  the (reduced) pre-collision
velocities in the  so-called inverse collision, resulting in the post-collision velocities ${\bf c}_{1}$ and
${\bf c}_{2}$ \cite{bp04}. The Heaviside function $\Theta (-{\bf c}_{12}\cdot {\bf e}\, )$ selects the
approaching particles and the factor $\chi$ equals the product of the Jacobian of the transformation
$\left({\bf c}_1^{\,\prime\prime}, \, {\bf c}_2^{\,\prime\prime}\right) \to \left({\bf c}_{1}, \, {\bf
c}_{2}\right)$ and the ratio of the lengths of the collision cylinders of the inverse and the direct
collisions \cite{bp04}:
\begin{equation}
\label{Jacobdef} \chi \equiv \frac{\left|{\bf c}_{12}^{\
\prime\prime}\right|}{\left|{\bf c}_{12}\right|} \, \frac{{\cal D}\left({\bf c}_1^{\
\prime\prime},{\bf c}_2^{\ \prime\prime} \right)} {{\cal D}\left({\bf c}_1,{\bf c}_2
\right)} \, .
\end{equation}

Now we substitute the Sonine-polynomial expansion for $\tilde{f}({\bf c},t)$ [Eq.~(\ref{Sdistrib})] into
Eq.~(\ref{sbe}) and neglect all terms with $p>3$. Multiplying the resulting equation with  $c_1^2$,
$c_1^4$ and $c_1^6$, and  integrating over ${\bf c}_1$, we obtain

\begin{eqnarray} \label{sys1}
\frac{\partial{\langle c^2\rangle}}{\partial t}+\frac{1}{T}\frac{dT}{dt}{\langle c^2\rangle}
&=& -\sqrt{\frac{2T}{m}} g_2(\sigma)\sigma^2n\mu_2+3\frac{m\xi^{2}_{0}}{2T} ,\\
\frac{\partial{\langle c^4\rangle}}{\partial t}+\frac{2}{T}\frac{dT}{dt}{\langle c^4\rangle} &=&
-\sqrt{\frac{2T}{m}} g_2(\sigma)\sigma^2n\mu_4+10{\langle c^2\rangle}\frac{m\xi^{2}_{0}}{2T},\nonumber \\
\frac{\partial{\langle c^6\rangle}}{\partial t}+\frac{3}{T}\frac{dT}{dt}{\langle c^6\rangle} &=&
-\sqrt{\frac{2T}{m}} g_2(\sigma)\sigma^2n\mu_6+24{\langle c^4\rangle}\frac{m\xi^{2}_{0}}{2T} \nonumber .
\end{eqnarray}
Here, \be \mu_p=-\int d{\bf c}c^{p}\tilde{I}(\tilde{f},\tilde{f}) \label{mup} \ee is
the $p$-th moment of the dimensionless collision integral. The moments $\mu_p$ can be
calculated analytically up to ${\cal O}(\delta^{10})$ using a formula manipulation
program as explained in detail in Ref. \cite{bp04}. They can be written as follows:
\begin{eqnarray}
\label{mupw}
 &&\mu_p=\sum_{k=0}^{20}\left( M^{(p,\,0)}_{k} + M^{(p,\,2)}_{k} a_2 + M^{(p,\,3)}_{k} a_3  +\right.\\
 &&+ \left.
M^{(p,\,22)}_{k} a_2^2 + M^{(p,\,33)}_{k} a_3^2 + M^{(p,\,23)}_{k}
a_2a_3\right)\delta^{k/2}\left(2u\right)^{k/20}   \nonumber
\end{eqnarray}
The numerical values of the coefficients $M^{(p,\,i)}_{k}$  for $\mu_p$ with $p=2, \,
4, \, 6$ are given in Appendix B.

Let us define the dimensionless time $\tau= t/\tau_c(0)$, where
\be
\tau_{c}^{-1}(t)=4\sqrt{\pi}g_{2}(\sigma)\sigma^{2}n\sqrt{\frac{T(t)}{m}}.
\label{tauc} \ee
Using  Eqs.~(\ref{cak1}) which express the the moments of the reduced velocity $\langle c^{2k} \rangle $ in
terms of the Sonine coefficients $a_k$,  we recast Eqs.~(\ref{sys1}), into the following form,
\begin{eqnarray}
\label{system1}
\frac{du}{d\tau} &=& -\frac{\sqrt{2}\mu_{2}}{6\sqrt{\pi}}u^{\frac{3}{2}}+\frac{K}{4}, \\
\frac{da_{2}}{d\tau} &=& \frac{\sqrt{2}\sqrt{u}}{3\sqrt{\pi}}\mu_{2}
\left(1+a_{2}\right)-\frac{\sqrt{2}}{15\sqrt{\pi}}\mu_{4}\sqrt{u}-\frac{Ka_2}{2u} , \nonumber \\
\frac{da_{3}}{d\tau} &=& \frac{\sqrt{u}}{\sqrt{2\pi}}\mu_{2}(1-a_{2}+a_{3})-   \nonumber
\\
&-&\frac{\sqrt{2}}{5\sqrt{\pi}}\mu_{4}\sqrt{u}+\frac{2\sqrt{2u}}{105\sqrt{\pi}}\mu_{6}-\frac{3a_3K}{4u}\,
, \nonumber
\end{eqnarray}
where the moments $\mu_p$, $p=2, \, 4, \, 6$ are the functions of the Sonine
coefficients $a_2$, $a_3$ and known  numerical constants $M^{(p,\,i)}_{k}$,
Eq.~(\ref{mupw}). The constant $K$ reads,
\begin{equation}
K = \frac{m^{3/2}\xi_0^2}{\sqrt{\pi}g_2(\sigma)\sigma^2nT_0^{3/2}} .
\end{equation}

Eqs.~(\ref{system1}) describe evolution of temperature and the velocity distribution
function for a heated  gas of viscoelastic particles. For $K = 0$, that is,  for
$\xi_{0}^{2}=0$, Eqs.~(\ref{system1}) reduce to corresponding equations of
Ref.~\cite{bb09}, which describe evolution of $a_2$ and $a_3$ in a force-free gas.

When a heated gas settles in a steady state, the  granular temperature and the velocity distribution
function become time-independent. Then Eqs.~(\ref{system1}) yield the algebraic equations for the
steady-state temperature and moments $\mu_p$, $p=2, \, 4, \, 6$:
\begin{eqnarray}
\label{consys1}
T_{\rm s.s.} &=& \left[\frac{3}{2\sqrt{2}}\frac{\xi^{2}_{0}m^{3/2}}{\mu_2g_2(\sigma)\sigma^2n}\right]^{2/3}, \\
\label{consys2}
\mu_4 &=& 5\mu_2, \\
\label{consys3}
\mu_6 &=& \frac{105}{4}\mu_2(1+a_2) .
\end{eqnarray}
Solving this system in the linear approximation with  respect to the dissipation
parameter $\delta$, we obtain  the steady state Sonine coefficients
\begin{equation}
\label{ss1} a_2^{\rm s.s.}=-A_2\delta, \hspace{0.7cm}  A_2=2^{1/5}\frac{157}{500}
\Gamma\left(\frac{21}{10} \right) C_1 \simeq 0.435 ,
\end{equation}
\begin{equation}
\label{ss2} a_3^{\rm s.s.} =-A_3\delta, \hspace{0.7cm}  A_3=2^{1/5}\frac{28}{500} \Gamma
\left(\frac{21}{10} \right) C_1 \simeq 0.078 .
\end{equation}
The above expressions for $a_2$ and $a_3$, together with Eq.~(\ref{mupw}) for $\mu_2$
yield the corresponding analytical expression for the steady state temperature $T_{\rm
s.s.}$.

\section{Effective restitution coefficient}
\label{Sepsilon}

\begin{figure}
\includegraphics[width=0.99\columnwidth]{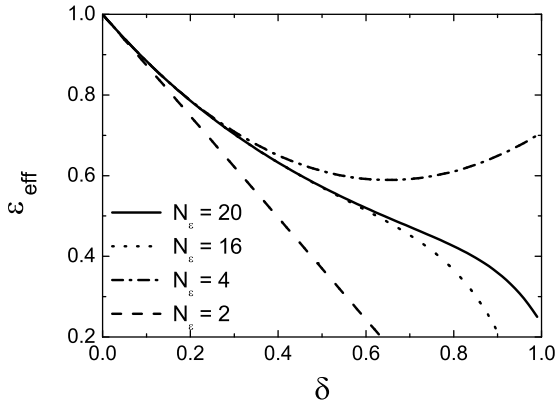}
\caption{An effective normal restitution coefficient $\varepsilon_{\rm eff}$, obtained as a "collision average" for $T=T_0$ and different values of $N_{\varepsilon}$ [Eq.~(\ref{epsav_int})].} \label{Gepsdelta}
\end{figure}

The expression (\ref{rerc}) for the restitution coefficient gives this quantity for a collision with a
particular impact velocity. A natural question arises, whether is it possible to define an average value of
$\varepsilon $, which characterizes the whole ensemble and may be exploited  as an effective constant
restitution coefficient $\varepsilon_{\rm eff}$ for all impacts; this however  should depend  on the current
temperature of a granular gas, see Eq.~(\ref{rerc}). Since $\varepsilon_{\rm eff}$ describes the collisions,
it is natural to define this as a "collision average", e.g. \cite{TrizackKrapPRL}:
\ba \label{colav} \varepsilon_{\rm eff} \! = \!\frac{ \int \!d {\bf v}_1 d {\bf v}_2 d {\bf e} f_1 f_2 \left|
{\bf v}_{12} \cdot {\bf e} \right| \Theta \left( -{\bf v}_{12} \cdot {\bf e}\right) \varepsilon \left( \left|
{\bf v}_{12} \cdot {\bf e} \right| \right) } {\int d {\bf v}_1 d {\bf v}_2  d {\bf e} f_1 f_2  \left| {\bf
v}_{12} \cdot {\bf e} \right| \Theta \left( -{\bf v}_{12} \cdot {\bf e}\right)  } , \nonumber   \ea
where the velocity distribution functions $f_1=f({\bf v}_1,t)$ and $f_2=f({\bf v}_2,t)$ are  given by Eq.
(\ref{Sdistrib}).  The integral in the above equation  can be calculated up to ${\cal O}(\delta^{10})$ using
a formula manipulation program as explained in detail in Ref. \cite{bp04}. Neglecting in the Sonine expansion
(\ref{Sdistrib}) for $f_1$ and $f_2$ the high-order coefficients $a_p$ with $p \ge 4$, we obtain,

\begin{eqnarray}
\label{epsav} &&\varepsilon_{\rm eff}(t)=1+\sum_{k=1}^{N_{\varepsilon}}\left[\delta^{k/2}\!\left(2u\right)^{k/20}\times\right.\\  \nonumber
&&\left.\times\frac{\!B_k\!+\!B_{k}^{(2)}\!a_2\!+\! B_{k}^{(3)}\!a_3\! +\! B_{k}^{(22)}\!a_2^2 \!+\!
B_{k}^{(33)}\!a_3^2 + B_{k}^{(23)}\!a_2a_3\!}{1+\!
B_{0}^{(2)}\!a_2\!+\!B_{0}^{(3)}\!a_3\!+\!B_{0}^{(22)}\!a_2^2\!+\!B_{0}^{(33)}\!a_3^2\!+\!B_{0}^{(23)}\!a_2a_3}\right]
\nonumber
\end{eqnarray}
with the numerical coefficients $B_k$ and $B_{k}^{(i)}$ given in the Appendix A. The above expression,
although general, is rather involved. If we neglect the dependence of $\varepsilon_{\rm eff}$ on the Sonine
coefficients (that is, perform the collision averaging of $\varepsilon$ with a Maxwellian distribution),
Eq.(\ref{epsav}) reduces to:
\be \label{epsav_int} \varepsilon_{\rm eff}(t) =1+ \sum_{k=1}^{N_{\varepsilon}}B_k  \, \delta^{k/2}
\left[2T(t)/T_0 \right]^{k/20} \, , \ee
where the above coefficients $B_k$ may be also written in a compact analytical form,
\begin{equation}
B_k=h_k\frac{2^{1+\frac{k}{20}}}{2+\frac{k}{10}} \Gamma \left[2+\frac{k}{20} \right]\,.
\end{equation}
Here $\Gamma(x)$ is the Gamma-function. We wish to stress again, that in opposite  to the impact-velocity
dependent coefficient $\varepsilon$, Eq.~(\ref{rerc}), the above restitution coefficient $\varepsilon_{\rm
eff}$ is the same for all collisions. Its dependence on the dissipation parameter $\delta$ is shown in
Fig.~\ref{Gepsdelta} for different $N_{\varepsilon}$.
As it follows from this figure the series (\ref{epsav_int}) demonstrates an excellent convergence for $\delta
\lesssim 0.5$.  In what follows we will use the number of terms in the expansion (\ref{epsav_int}) equal to
$N_{\varepsilon}=20$, which guarantees an accurate description of the effective restitution coefficient
$\varepsilon_{\rm eff}$ for the dissipative parameter $\delta$ up to $\delta \lesssim 0.5$, that is, for a
rather large dissipation.

The system of equations, Eqs.~(\ref{system1}) may be also used for a constant
restitution coefficient, or, equally for the quasi-constant coefficient
$\varepsilon_{\rm eff}$, Eq.~(\ref{epsav}), provided the respective moments $\mu_p$ are
expressed in terms of $a_2$, $a_3$ and $\varepsilon_{\rm eff}$, as it follows for the
case of a constant $\varepsilon$ \cite{bp06}. We write these in the following form,

\begin{eqnarray}
\label{muconst}
&&\mu_2=\frac{1}{16384}\sqrt{2\pi}(1-\varepsilon_{\rm eff}^2)(35a_3^2+144a_2^2\\
\nonumber &&+120a_2a_3+3072a_2+256a_3+16384)\\ \nonumber &&\mu_4 =
\frac{1}{32768}\sqrt{2\pi}(1+\varepsilon_{\rm eff})\sum_{k=0}^{3}\left( N^{(4,\,0)}_{k}
+ N^{(4,\,2)}_{k} a_2  \right.\\\nonumber &&+\left. N^{(4,\,3)}_{k} a_3 +
N^{(4,\,22)}_{k} a_2^2 + N^{(4,\,33)}_{k} a_3^2 +
N^{(4,\,23)}_{k}a_2a_3\right)\varepsilon_{\rm eff}^k\\\nonumber
&&\mu_6=-\frac{3}{262144}\sqrt{2\pi}(1+\varepsilon_{\rm eff})\sum_{k=0}^{5}\left(
N^{(6,\,0)}_{k} + N^{(6,\,2)}_{k} a_2 \right.\\\nonumber &&+\left.
N^{(6,\,3)}_{k} a_3  +N^{(6,\,22)}_{k} a_2^2 + N^{(6,\,33)}_{k} a_3^2 +
N^{(6,\,23)}_{k}a_2a_3\right)\varepsilon_{\rm eff}^k 
\end{eqnarray}
with the coefficients $N^{(p,\,i)}_{k}$ for $p=4,6$  listed in the Appendix B.

\begin{figure}
\includegraphics[width=0.99\columnwidth]{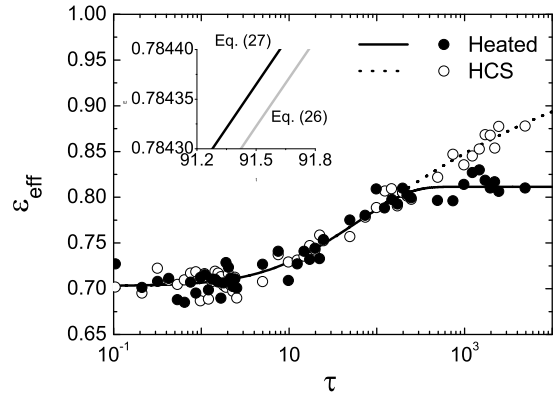}
\caption{Evolution of the average normal restitution coefficient, obtained in the MD simulations (symbols) and of the effective "quasi-constant" restitution coefficient $\varepsilon_{\rm eff}$, given by Eq.~(\ref{epsav}) (grey lines) and Eq.~(\ref{epsav_int}) (black lines). The difference between solutions of Eq.~(\ref{epsav}) and Eq.~(\ref{epsav_int}) is shown at the inset (for heated case) and is not visible on the main plot. The dissipation constant $\delta=0.3$. The dimensionless time $\tau=t/\tau_{c}(0)$ is expressed in terms of the initial mean collision time $\tau_c(0)$, Eq.~(\ref{tauc}). Note that the restitution coefficient tends to unity in the homogeneous cooling state and to a steady-state value for the heated gas. } \label{Gepstau}
\end{figure}

Using Eqs.~(\ref{system1}) and Eqs.~(\ref{muconst}) together with Eq.~(\ref{epsav}) one can find temperature
of a gas and the Sonine coefficients $a_2$ and $a_3$. Ultimately, $\varepsilon_{\rm eff}$ as a function of
current temperature, as it follows from Eq.~(\ref{epsav}),  may be  found. This,  however,  is very close to
the simplified form (\ref{epsav_int}), so that the difference between the results, obtained with the use of
the complete expression (\ref{epsav}) and the simplified one, (\ref{epsav_int}), can be hardly distinguished on the figures (See Fig.~\ref{Gepstau}). Hence, for practical purposes one can safely use the simple expression (\ref{epsav_int}) for the effective quasi-constant restitution coefficient.

\section{Event-driven simulations}
\label{sec2}

\subsection{Force-free Gas}
\label{sec2.1}

We perform event-driven  MD simulations \cite{Compbook} of a granular gas of smooth
viscoelastic particles with the restitution coefficient (\ref{rerc}).  The grains are
modeled as identical hard spheres of mass $m = 1$ and diameter $\sigma = 1$. In
event-driven simulations particles move freely between pairwise collisions, where the
velocities of the grains are updated according to Eq.~(\ref{v1v2}). We use the system
of $N = 2.048\times10^6$ particles placed in a three-dimensional cubic box with the
edge $L=418$, which corresponds to the number density of $n=0.028$. This number density is rather small, which savely allows to approximate the contact value of the correlation function by unity, $g_2(\sigma)\simeq 1$. Periodic boundary conditions are applied in all three directions.

The system was initialized by placing randomly particles  in the box. For initial
particles' velocities  randomly directed vectors  of the same length have been
assigned. The initial total momentum $\sum_i m{\bf v}_i(0)$ was in this case with a
high accuracy zero. We start with $\varepsilon = 1$, which corresponds to elastic
collisions, and allow the system to relax to a Maxwellian velocity distribution, which
used to happen in a few collisions per particle. Hence, a homogeneous system with a
Maxwellian distribution was an initial condition for each MD run; then we  evolve the
system with a velocity-dependent restitution coefficient (\ref{rerc}). In simulations
we used the initial granular temperature $T(0) = 400/3$ with the corresponding initial
mean collision time $\tau_c(0) = 0.436$. All statistical quantities presented here have
been obtained as averages over $25$ independent runs. We have confirmed that the system
remains approximately homogeneous during our simulations and that the Molecular Chaos
hypothesis holds true  with a high degree of accuracy, as it used to be in dilute
gases, e.g. \cite{brey2004}.

In Fig.~\ref{Gepstau}, we plot time dependence  of the current  average of the restitution coefficient; this
was obtained, averaging $\varepsilon$ over $(N/20)$ successive particles collisions. As it follows from the
figure, the average restitution coefficient is in a good agreement with the theoretical prediction for the
effective "quasi-constant" coefficient $\varepsilon_{\rm eff}$,  Eq.~(\ref{epsav_int}), for both, freely evolving and heated granular gases.

In Fig.~\ref{Guinset}, we present simulation results for the time dependence of the
reduced granular temperature $u(\tau)$ on the dimensionless time $\tau=t/\tau_{c}(0)$,
expressed in terms of the initial mean collision time $\tau_c(0)$, Eq.~(\ref{tauc}).
In a HCS  temperature decays asymptotically as the  power-law $u(\tau)\sim \tau^{-5/3}$
\cite{bp04,bp00,bb09} while in a heated gas it relaxes to a steady-state value.

Next, we analyze evolution of the velocity distribution function, characterized by the
time-dependent Sonine coefficients  [cf. Eq.~(\ref{Sdistrib})]. In order to calculate
$a_2$ and $a_3$ numerically, we compute the moments of the velocity distribution
function $\langle c^{2k}\rangle$ and use Eqs. (\ref{cak1}).
In Fig.~\ref{Ga2a3d015}, we show evolution of $a_2$ and $a_3$ for a force-free gas for $\delta = 0.15$, while in Fig.~\ref{Ga2a3a4a5} the according simulation data for the Sonine coefficients $a_2,\,a_3,\,a_4$ and $a_5$ for $\delta=0.1$ and $\delta =0.3 $ are given.

\subsection{Uniformly heated gas}
\label{sec2.2}

The equation of motion for $i$-th particle  in a white-noise thermostat, that is, for
the gas under a uniform heating, reads,

\be
m\frac{d{\bf v}_{i}}{dt}={\bf F}^{\rm coll}_{i}+{\bf F}^{\rm ext}_{i} \, .
\ee
Here ${\bf F}^{\rm coll}_{i}$ is the force acting on $i$-th particle only in a pairwise
collision with another gas particle, while ${\bf F}^{\rm ext}_{i}$ is the external
force acting from the thermostat. Since we use the event-driven simulations
\cite{Compbook}, we do not need to solve explicitly the equation of motion for
colliding particles with the forces ${\bf F}^{\rm coll}$. Instead, we directly apply
the collision rules (\ref{v1v2}) and write the after-collision velocities, expressed in
terms of the pre-collision ones with  the  restitution coefficient $\varepsilon$,
Eq.~(\ref{rerc}) \cite{bp04,Compbook}. For the external force ${\bf F}^{\rm ext}$ we
use the model of a Gaussian white noise:
\be \langle{\bf F}^{\rm ext}_{i}(t)\rangle=0 , \quad \langle F^{\rm ext}_{i,\alpha}(t)
F^{\rm ext}_{j,\beta}(t')\rangle=m^{2}\xi^{2}_{0}\delta_{ij} \delta_{\alpha
\beta}\delta(t-t')\, , \ee
where $\alpha, \beta = x,y,z$, and $\xi^{2}_{0}$ characterizes the magnitude of the
stochastic force.

\begin{figure}
\includegraphics[width=0.99\columnwidth]{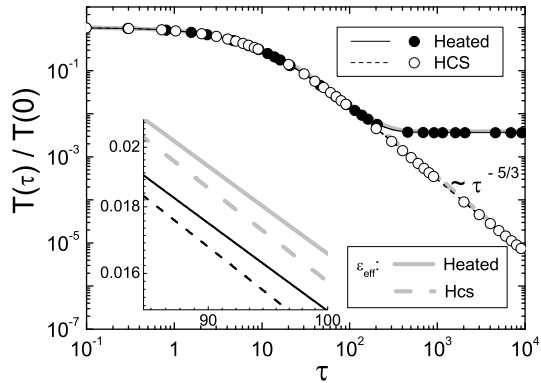}
\caption{Evolution of the reduced granular temperature $u(\tau)=T(\tau)/T(0)$ for the force-free and heated
granular gases. The dissipation parameter $\delta = 0.3$. The symbols denote simulation results, thin black lines -- theory (numerical solution of Eqs. (\ref{system1}), where $\mu_p$ are given by Eqs. (\ref{mupw})). 
For $\tau \gg 1$ $T/T(0)\sim\tau^{-5/3}$ for a force-free gas and $T\rightarrow T_{s.s.}$ for a heated gas. Thick grey  lines depict the theoretical predictions for temperature of a granular gas with a "quasi-constant"
restitution coefficient $\varepsilon_{\rm eff}(t)$ (numerical solution of Eqs. (\ref{system1}), where $\mu_p$
are given by Eqs. (\ref{muconst}) and $\varepsilon_{\rm eff}(t)$ - by Eq. (\ref{epsav_int})).} \label{Guinset}
\end{figure}

\begin{figure}
\includegraphics[width = 0.99\columnwidth]{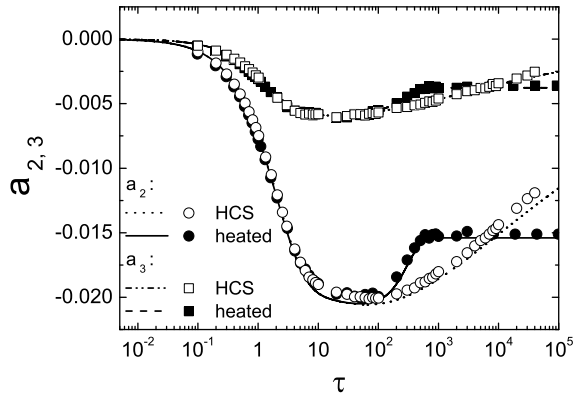}
\caption{Evolution of $a_2$ and $a_3$ for the force-free and heated granular gases with
$\delta = 0.15$. The symbols correspond to simulation results, the lines to our theory
(numerical solution of Eqs.~(\ref{system1}), where $\mu_p$ are given by Eqs.
(\ref{mupw})).} \label{Ga2a3d015}
\end{figure}

\begin{figure*}
\includegraphics[width = 0.99\columnwidth]{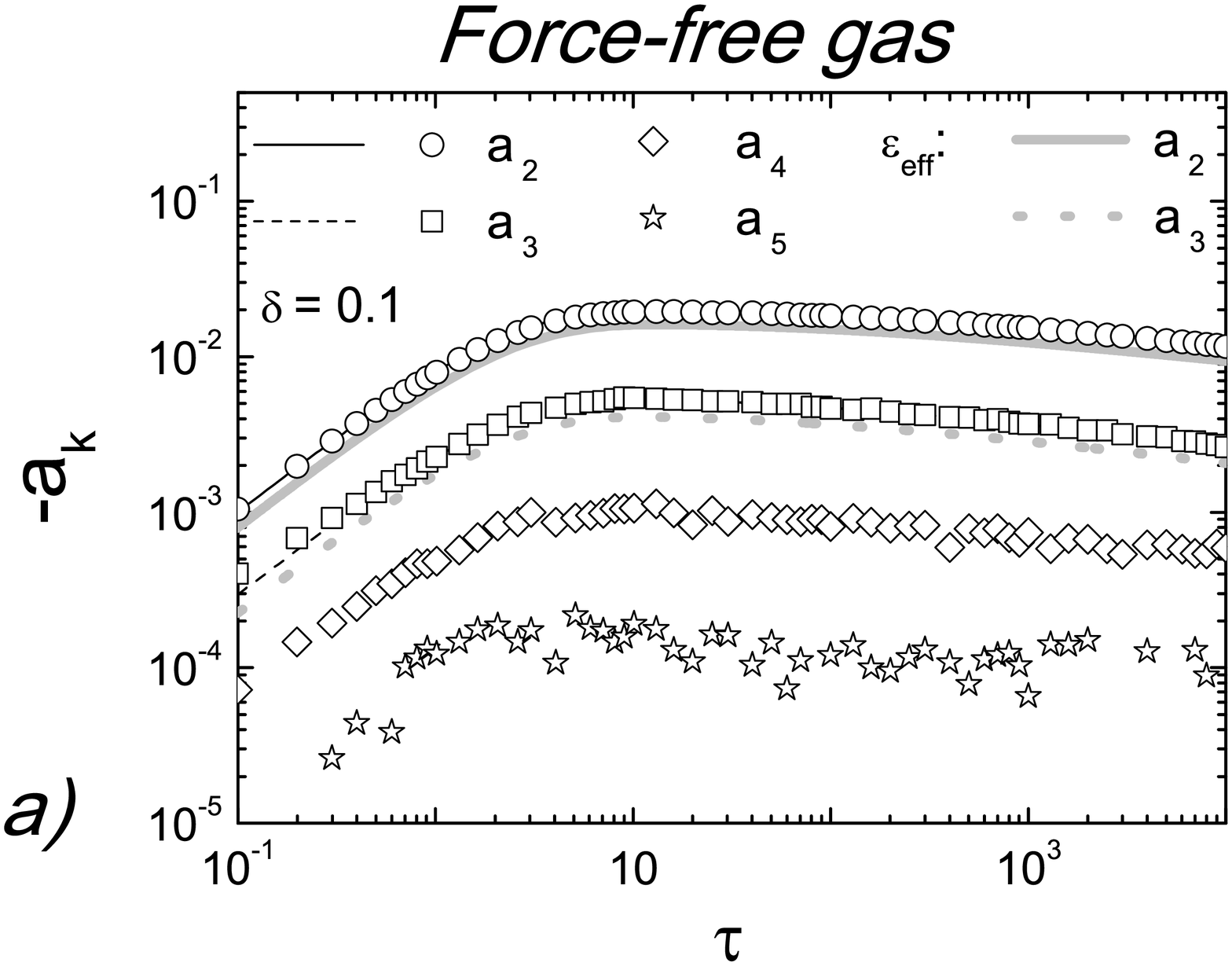}
\includegraphics[width=0.99\columnwidth]{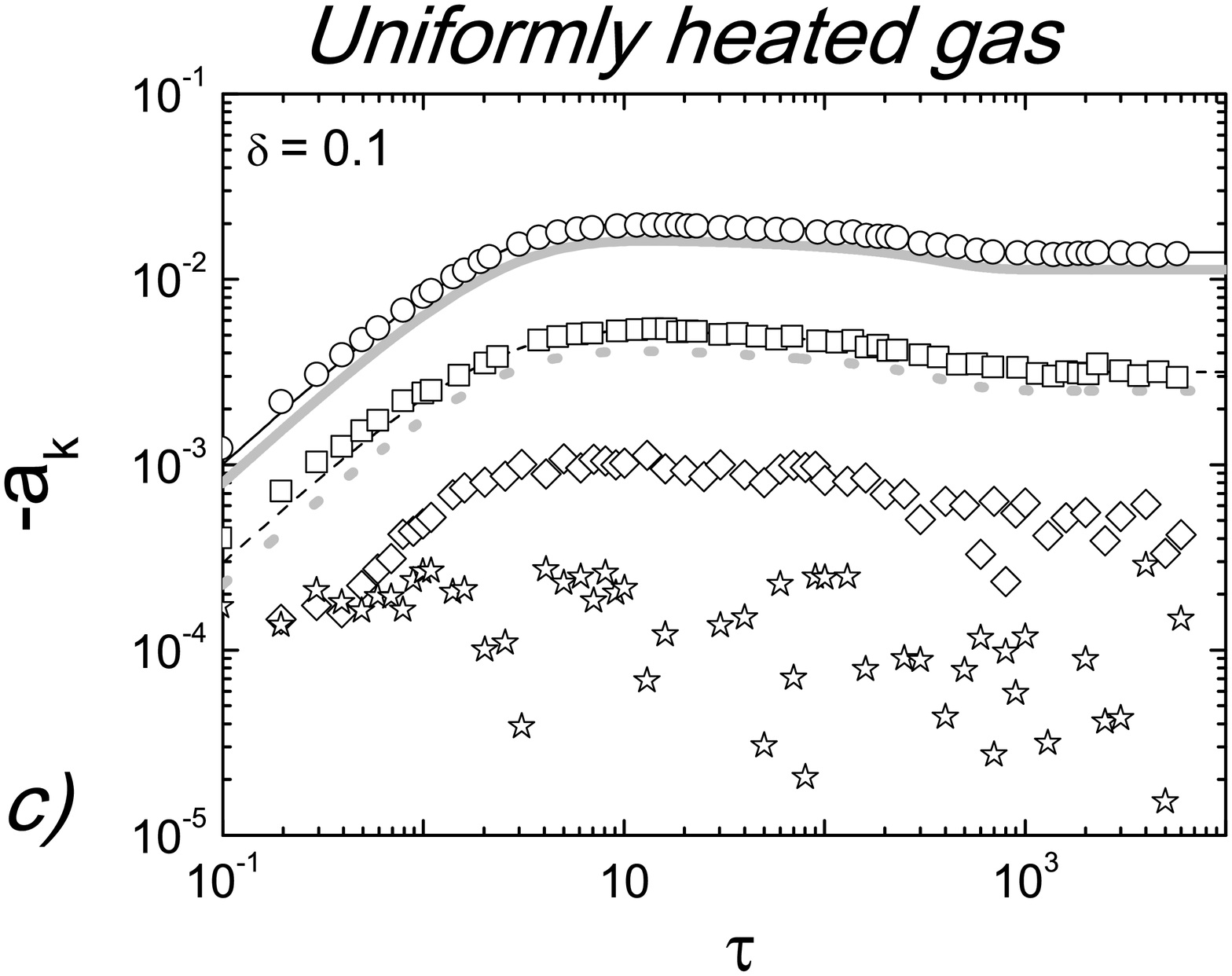}
\includegraphics[width=0.99\columnwidth]{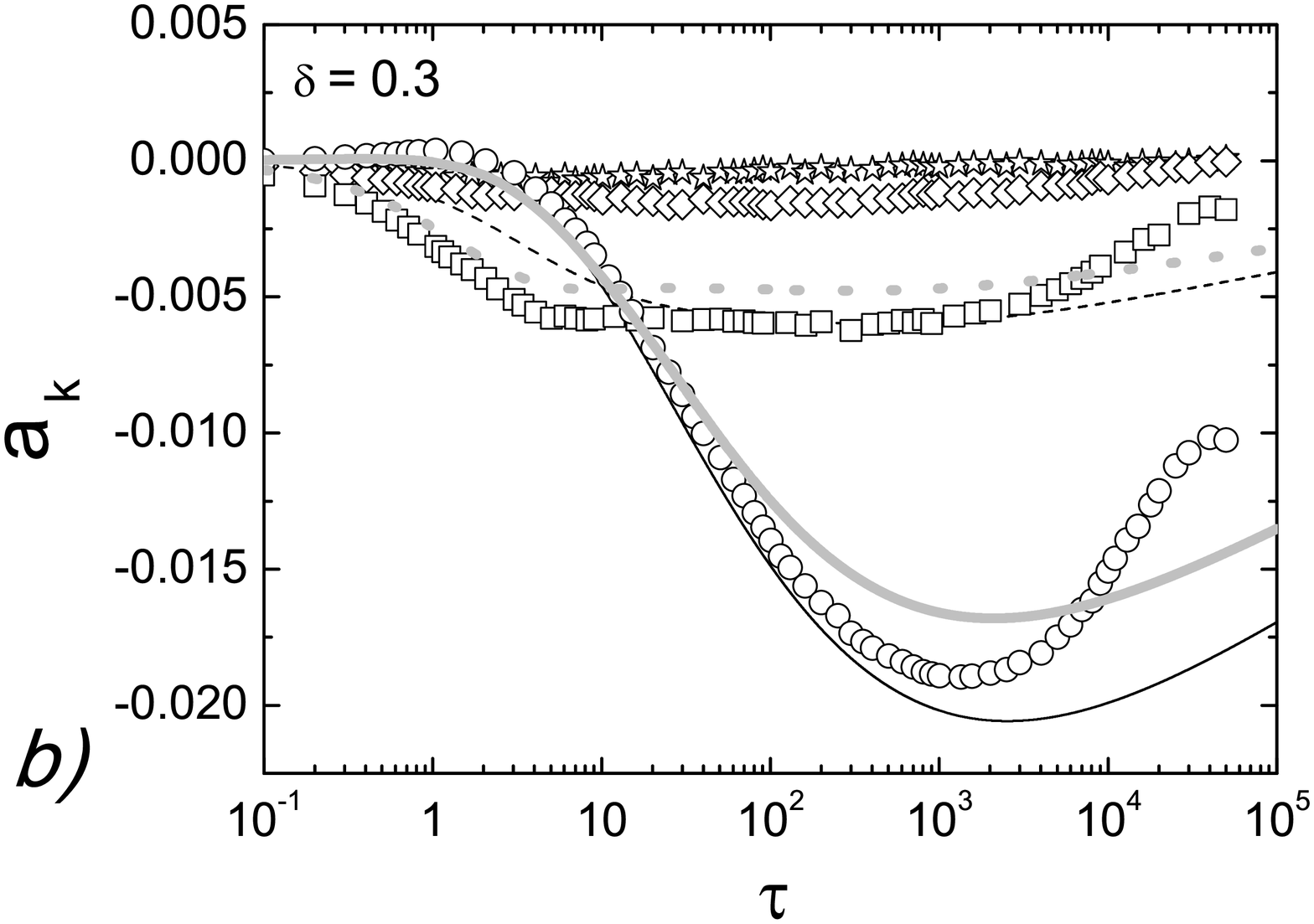}
\includegraphics[width=0.99\columnwidth]{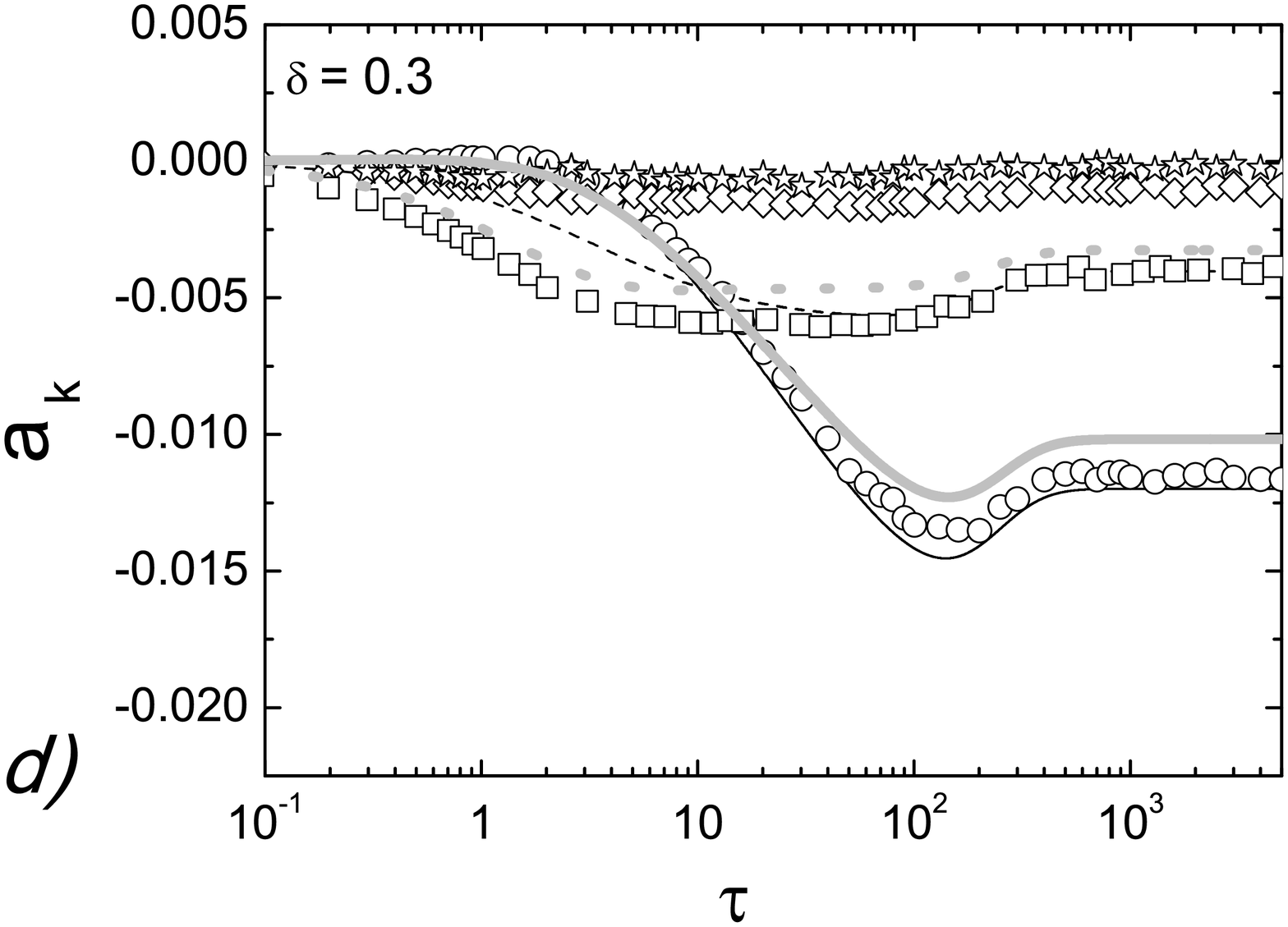}
\caption{Evolution of the Sonine coefficients $a_k$ ($k=2,3,4,5$) for a force-free (a,b) and uniformely heated (c, d) granular gas with $\delta = 0.1$ (a, c) and $\delta = 0.3$ (b,d). The symbols correspond to the MD simulation results and the thin black lines - to the theory (numerical solution of Eqs.~(\ref{system1}), where $\mu_p$ are given by Eqs. (\ref{mupw})). Thick grey lines depict theoretical predictions (numerical solution of Eqs.~(\ref{system1}), where $\mu_p$ are given by Eqs. (\ref{muconst})) for a
granular gas with a "quasi-constant" restitution coefficient $\varepsilon_{\rm eff}(t)$, Eq.~(\ref{epsav_int}) .} \label{Ga2a3a4a5}
\end{figure*}

We apply the  algorithm suggested in Ref.~\cite{wm} (see also
\cite{Pag1999,Sant2000}):~During the event-driven simulations, we heat the system after
a time-step $dt$ by adding to the velocity of each $i$-th particle a random increment,
which mimics the heating by noise,

\be
v_{i,\alpha}(t+dt)=v_{i, \alpha}(t)+\sqrt{r}\sqrt{dt}\varphi \, ,
\label{he}
\ee
where $\alpha = x,y,z$ and  $\varphi $ is a random number uniformly distributed within
the interval $\left[-0.5, \, 0.5\right]$, and $r$ is the amplitude of the noise,  $r =
12\xi_0^2$. After the change in velocities, the system is transferred to the
centre-of-mass frame,
\be {\bf v}_{i}={\bf v}_{i}-\frac{1}{N}\sum^{N}_{i=1}{\bf v}_{i}
\,  , \nonumber
\ee
to ensure the conservation of  the momentum. In simulations we used $r = 0.1$ and $dt =
0.1$, which implies that the random kicking interval is small as compared to the mean
free time. The number of particles and the system size were the same as in the
force-free simulations; the initial conditions have been also prepared as in the
force-free case. All statistical quantities were obtained as averages
 over $10$ independent runs. The results are presented at Figs.~\ref{Gepstau}-\ref{Ga2a3a4a5}, where we show evolution of the average restitution coefficient (Fig.~\ref{Gepstau}),  of temperature (Fig.~\ref{Guinset}), and  of the Sonine coefficients (Figs.~\ref{Ga2a3d015} and \ref{Ga2a3a4a5}).
 
 \section{Results and discussion}
\label{sec4}

Let us compare the results of the MD simulations with the predictions of our theory. As
we already mentioned, the current average of the restitution coefficient, obtained by a
straightforward averaging of $\varepsilon$ in successive collisions is in a good
agreement with the theoretical value of the effective quasi-constant restitution
coefficient $\varepsilon_{\rm eff}$, found as a collision average from the kinetic
theory, Fig.~\ref{Gepstau}. In Fig.~\ref{Guinset} evolution of granular temperature is
shown. As it follows from the figure,  during the early stages of cooling, the
temperature of a heated granular gas evolves in the same way as the respective
force-free gas with the same $\delta$. However, while the temperature of the heated gas
settles at  a steady-state value, the force-free gas continues to cool down. The
results of our  theory (numerical solution of system of differential equations (\ref{system1}), where moments of collisional integral $\mu_p$ are given by Eqs. (\ref{mupw})) are in an excellent agreement with the simulation data for both the homogeneous cooling state and  uniformly heated gas. Moreover, the kinetic theory
based on the effective quasi-constant restitution coefficient $\varepsilon_{\rm eff}$ (numerical solution of Eqs. (\ref{system1}), where $\mu_p$ are given by Eqs. (\ref{muconst}))
yields almost the same accuracy for the description of granular temperature as the
complete theory.

In Fig.~\ref{Ga2a3d015}, evolution of the Sonine coefficients $a_2$ and $a_3$ is
presented for the force-free and heated gas with $\delta = 0.15$.  At the initial stage
of evolution, for $\tau \leq 100$, the values of $a_2$ and $a_3$ in a heated and
force-free gas are practically indistinguishable. At later times, the Sonine
coefficients for the heated gas settle at the steady-state
[cf.~Eqs.~(\ref{ss1})-(\ref{ss2})]. For the freely evolving gas they continue to decay
and evolve as following:~Starting from zero,  $a_k(0)  = 0$, ($k=1,2, \ldots$), which
corresponds to the initial Maxwellian distribution, $a_2$ and  $a_3$ decrease, reach a
minimum, and eventually return back to zero, see Fig.~\ref{Ga2a3d015}. As the thermal
velocity decreases during the gas cooling, the behavior of the system tends to that of
a gas of elastic particles, that is, the velocity distribution tends asymptotically to
a Maxwellian. For both, the force-free and heated gases, our theory perfectly agrees
with the computer simulations.

To investigate the  convergence of the Sonine-polynomial expansion (\ref{Sdistrib}) we
analyze the high-order Sonine coefficients, $a_4$ and $a_5$. In  Fig.~\ref{Ga2a3a4a5} (a) the
simulation data for the time dependence of $a_2,\,a_3,\,a_4,\, a_5$  along with the
analytical predictions for $a_2$ and $a_3$ are shown. As it may be
seen from Fig.~\ref{Ga2a3a4a5}(a), $a_4$ and $a_5$ evolve in a similar fashion as $a_2$ and
$a_3$, that is, their magnitudes first increase, reach maximum and relax back to zero.
Certainly, this is not surprising, since it is a simple manifestation of the fact that
the system starts from a  Maxwellian velocity distribution due to initial conditions
and returns to it later, when the collisions become elastic. Furthermore, $a_4$ is an
order of magnitude smaller than $a_3$, while $a_5$ is $\sim 10^2$ times smaller than
$a_2$. This confirms the convergence of the Sonine-polynomial expansion for the
velocity distribution, at least for these values of $\delta$. In this figure we also
show, in addition to the predictions of the complete theory, theoretical results for
the case of an effective quasi-constant restitution coefficient $\varepsilon_{\rm eff}(t) $. Again we see that the accuracy of the simplified theory, based on the
effective $\varepsilon_{\rm eff}(t) $,  is surprisingly good.

In Fig.~\ref{Ga2a3a4a5}(b) evolution of the Sonine coefficients  for the larger dissipation,
$\delta =0.3$ is illustrated. The behavior of the coefficients is qualitatively the
same as for the smaller $\delta$. Moreover, the subsequent coefficients $a_k$ decrease
by an order of magnitude with increasing  $k$, which again indicates the convergence of
the Sonine-polynomial expansion for this dissipation. At the same time a rather
noticeable discrepancy between the kinetic theory and the MD results is obvious. In the
earlier work \cite{bb09}, two of us have reported that the expansion (\ref{rerc}) for
$\varepsilon$ with $N_{\varepsilon}=20$  does not yield an acceptable accuracy for the
Sonine coefficient $a_2$ and $a_3$ for $\delta \geq 0.3$ (when the complete theory is
used). The discrepancy  between the theoretical curves and the simulation data, seen in
Fig.~\ref{Ga2a3a4a5}(b),  seemingly supports this theoretical conclusion.

Figure~\ref{Ga2a3a4a5}(c) demonstrates the convergence of the Sonine-polynomial expansion
for heated granular gases. As in Fig.~\ref{Ga2a3a4a5}(a), we plot here  the dependence of
$a_2,\,a_3,\,a_4$ and $a_5$ on $\tau$ for $\delta=0.1$. It may be seen  that the
successive Sonine coefficients decrease by an order of magnitude, which clearly
indicates the convergence of this series. As it has been already mentioned in the
discussion of Fig.~\ref{Ga2a3d015}, our complete theory is in a perfect agreement with
the simulation data for the heated gas, but here we also notice  that  the simplified
model with a quasi-constant restitution coefficient $\varepsilon_{\rm eff}$ gives rather
accurate description of the evolution of $a_2$ and $a_3$ too.

Finally, in Fig.~\ref{Ga2a3a4a5}(d), we plot time dependence of $a_2,\,a_3,\,a_4$ and $a_5$
for a heated gas with relatively large dissipation, $\delta=0.3$. In this case one can
again see a noticeable discrepancy between our theory and the numerical results (cf.
Fig.~\ref{Ga2a3a4a5}(b) for the freely-evolving gas). Nevertheless, smallness of the
high-order coefficients $a_4$ and $a_5$ seemingly manifests the  convergence of the
Sonine series. Interestingly, although the simulation data deviate from the theoretical
curves during the relaxation to the steady state, in the steady state itself, the
numerical and theoretical values of the complete theory for the Sonine coefficients
$a_2$ and $a_3$ are very close, see Fig.~\ref{Ga2a3a4a5}(d) for $\tau > 10^3$. At the same
time the results of the simplified model of the quasi-constant restitution coefficient
$\varepsilon_{\rm eff}$ do not agree well with the MD results at the steady state for
this dissipation.


\section{Conclusion}
\label{sec5}

To date, most studies of granular gases have been focused on systems with a constant
restitution coefficient, although the basic physics requires impact-velocity dependence
of $\varepsilon$. Here we studied numerically, by means of Molecular Dynamics, and
theoretically a granular gas of particles with the impact-velocity dependent
restitution coefficient, as it follows from the simplest first-principle model of
viscoelastic spheres. Since the impact-velocity  dependence of $\varepsilon$
significantly complicates the description of a granular gas dynamics, we tried to
develop a simplified collision model. Namely, we considered a model where $\varepsilon$
is the same for all collisions, but depends on the current value of the thermal
velocity of the gas, or, equally, on granular temperature.  Using the  collision
average of the kinetic theory, we obtained an explicit expression for this
quasi-constant restitution coefficient $\varepsilon_{\rm eff} [T(t)]$.

We explored  evolution of granular temperature and the velocity distribution function
in large-scale MD simulations of force-free and uniformly heated gases and developed a
theory of the velocity distribution function $f({\bf v},t)$ for the case of driven
gases. In the  MD simulations we found a few first Sonine coefficients $a_2$, $a_3$,
$a_4$, $a_5$, which quantify deviations of  $f({\bf v},t)$  from a Maxwellian
distribution and noticed that the successive-order Sonine coefficients  decrease by an
order of magnitude with respect to the preceding ones. In other words, we observed that
$a_3$ is approximately $10$ times smaller than $a_2$, $a_4$ is $10$ times smaller than
$a_3$ and $a_5$ is $10$ time smaller than $a_4$. This implies the convergence of the
Sonine series for both heated and force-free gases in the range of dissipation studied.

We obtained theoretical predictions for $a_2$ and $a_3$ using the complete theory and
the one based on an effective quasi-constant restitution coefficient  $\varepsilon_{\rm eff}$. We found that the results  of our complete theory for granular temperature are in
a very good agreement with the simulation data for both, force-free and heated gases.
Furthermore, if dissipation is not large, $\delta =0.1$ or $\delta=0.15$, the
dependence on time of the Sonine coefficient, obtained in simulations is also in an
excellent agreement with the theoretical predictions of the complete theory for  a
homogeneous cooling state (previously existed theory) as well as for a uniform heating
(the novel theory). At the same time for large dissipation, $\delta =0.3$, a noticeable
discrepancy between the theoretical and numerical curves is observed, especially for
the homogeneous cooling; deviations for the uniform heating are also noticeable, but
less pronounced. Moreover the steady-state values of $a_2$ and $a_3$ are given rather
accurately by the new theory for  heated gases.

We also analyzed the accuracy of the  simplified model based on the effective
quasi-constant restitution coefficient $\varepsilon_{\rm eff} (t)$. We have observed
that this model can accurately describe evolution of temperature for all studied
dissipations and of the velocity distribution function, provided dissipation is not
large, for both, force-free and heated gases. However, for large dissipation the
simplified model allows only qualitative description of $f({\bf v},t)$ and fails to
provide a quantitative one. Still we wish to stress that the simple model, where
$\varepsilon_{\rm eff} (t)$ is determined by the granular temperature and material
properties of the grains may be important for applications. Indeed, the description of
a granular gas dynamics for the case of a constant restitution coefficient  is
significantly simpler than that of an impact-velocity dependent $\varepsilon$.

\begin{acknowledgments}
A.B. gratefully acknowledges the financial support of this work by Russian Foundation for Basic Research (RFBR, project 12-02-31351) and the use of the facilities of the Chebyshev supercomputer of Moscow State University.
\end{acknowledgments}
\appendix*
\section{Numerical coefficients}

The effective restitution coefficient $\varepsilon_{\rm eff}$ and moments of the dimensionless collision
integral $\mu_p=-\int d{\bf c}c^{p}\tilde{I}(\tilde{f},\tilde{f})$, where $p=2, \, 4, \, 6$, can be
calculated using a formula manipulation program as explained in detail in Ref. \cite{bp04}. Below we present
coefficients $B_k^{(i)}$ which define $\varepsilon_{\rm eff}$ [Eqs.~(\ref{epsav_int} and (\ref{epsav})], numerical values of the coefficients $M_{k}^{(p,i)}$ for $\mu_p$ ($p=2, 4, 6$) [Eqs. (\ref{mupw}] in the case of a granular gas of viscoelastic particles and the coefficients $N_{k}^{(p,i)}$ for $\mu_p$ ($p= 4, 6$) [Eqs.
(\ref{muconst})] for a granular gas with a constant restitution coefficient.

\begin{table}[h]
\caption{\label{Table0} Numerical coefficients $B_k$ and $B_k^{(i)}$ for calculation of the effective
quasi-constant restitution coefficient $\varepsilon_{\rm eff}$ [Eq. (\ref{epsav})-(\ref{epsav_int})].}
\begin{ruledtabular}
\begin{tabular}{ | l | l | l | l | l | l | l |}
\hline $k$&$B_k$ & $B_k^{(2)}$ & $B_{k}^{(3)}$ & $B_{k}^{(22)}$&$B_{k}^{(33)}$&$B_{k}^{(23)}$
\\ \hline 0& - &-0.0625& -0.0156 & -0.0146   &  -0.0064  & 0.017  \\
\hline 1& 0 &0& 0 & 0   &  0  & 0  \\
\hline 2&-1.176&0.07&0.016&0.0148&0.00616&0.0168\\
\hline 3& 0 &0& 0 & 0   &  0  & 0  \\
\hline 4& 0.84 & -0.044   &-0.0096 &-0.00826 &-0.00326 & -0.009\\
\hline 5& 0.287&-0.0135 & -0.0028 &-0.00237 & -0.00091&-0.0026\\
\hline 6   &-0.578 &0.0231 &0.0046&0.0038 & 0.00142 &0.0041\\
\hline 7& -0.524 &0.0167 &  0.0032& 0.00258  &  0.000937 &0.0027 \\
\hline 8& 0.408&-0.00919 &-0.00168 &-0.00133 &  -0.00047 &-0.0014\\
\hline 9 & 0.547  &-0.0065 & -0.0011 & -0.00087 &-0.0003 &-0.00089\\
\hline 10&-0.184 & 0&0 &0 &0 & 0  \\
\hline 11& -0.479 &-0.0063&-0.00099 & -0.000727 &-0.000235 &-0.00072 \\
\hline 12&  -0.0586  &-0.00161 &-0.00024 &-0.00017 &-0.000054 &-0.00017\\
\hline 13& 0.398 &0.017& 0.00243 &0.00169 &0.00051 &0.0016 \\
\hline 14& 0.218 &0.013 & 0.00175 &0.00118 &0.00035&0.0011 \\
\hline 15& -0.277 &-0.022 & -0.0027&-0.00178 &-0.00051 &-0.0016 \\
\hline 16& -0.289 &-0.028 &-0.00329 &-0.002 &-0.00058 &-0.0019 \\
\hline 17& 0.112 &0.0133 & 0.0014 &0.00089 &0.00024 &0.00079\\
\hline 18&0.308  &0.043 & 0.0043 &0.00258&0.00067 &0.0022 \\
\hline 19& 0.0478 &0.0078 &0.00071  & 0.00042&0.0001&0.00035 \\
\hline 20& -0.276 &-0.0517 &-0.0043 &-0.0024 &-0.00059 &-0.002 \\
\end{tabular}
\end{ruledtabular}
\end{table}

\newpage

\begin{table}[h]
\caption{\label{Table1} Numerical coefficients for $\mu_2$ for a gas of viscoelastic particles [Eqs.
~(\ref{mupw})].}
\begin{ruledtabular}
\begin{tabular}{ | l | l | l | l | l | l | l |}
     \hline
     $k$ & $M^{(2,\,0)}_{k}$ & $M^{(2,\,2)}_{k}$ & $M^{(2,\,3)}_{k}$ & $M^{(2,\,22)}_{k}$& $M^{(2,\,33)}_{k}$& $M^{(2,\,23)}_{k}$ \\ \hline
     0& 0&  0&  0   &0& 0   &0\\ \hline
     1& 0&  0&  0   &0& 0   &0\\ \hline
     2& 6.49&    1.56& 0.10&    0.054& 0.012&    0.044  \\ \hline
     3& 0&  0&  0   &0& 0   &0\\ \hline
     4& -9.29&   -2.76&   -0.14&   -0.067&   -0.014&   -0.052\\ \hline
5&  -1.80&   -0.59&   -0.025    &-0.012&  -0.0023&   -0.0087\\
\hline 6&  10.40&    3.74&    0.12 &0.056& 0.011
&0.041\\ \hline 7& 5.91&    2.32& 0.058 &0.025&
0.0047&    0.018\\ \hline 8& -10.44&   -4.46    &-0.074&
-0.031&   -0.0055&   -0.021\\ \hline 9& -10.53&   -4.88&
-0.04&-0.016&  -0.0028&   -0.011\\ \hline 10& 8.77&
4.39&    -6.28E-08&   1.57E-07& -6.91E-07 &-1.10E-06\\ \hline 11&
14.13&    7.60&-0.063&  -0.023&   -0.0036&
-0.015\\ \hline 12& -4.54&   -2.62&   0.044&    0.015
&0.0023&    0.0093\\ \hline 13& -16.37&    -10.12&   0.25&
0.080 &0.012&   0.050\\ \hline 14& -1.58&    -1.04&
0.035 &0.010&   0.0015 &0.0063\\ \hline 15& 16.74&
11.77&    -0.49    &-0.14&  -0.018&   -0.08\\ \hline
16& 8.09&    6.05&    -0.30    &-0.079&  -0.01&   -0.045\\
\hline 17& -14.29&   -11.33&   0.66 &0.16    &0.02&   0.089\\
\hline 18& -13.91&   -11.69&   0.779 &0.175    &0.02&   0.0935\\
\hline 19& 8.82&    7.83&    -0.59&   -0.12 &-0.013&  -0.063\\
\hline 20& 18.13&    16.99&    -1.42&   -0.27& -0.028& -0.13\\ \hline
\end{tabular}
\end{ruledtabular}
\end{table}

\newpage

\begin{table}[h]
\caption{\label{Table2} Numerical coefficients for $\mu_4$ for a gas of viscoelastic particles [Eqs.
~(\ref{mupw})].}
\begin{ruledtabular}
\begin{tabular}{ | l | l | l | l | l | l | l |}
     \hline
     $k$ & $M^{(4,\,0)}_{k}$ & $M^{(4,\,2)}_{k}$ & $M^{(4,\,3)}_{k}$ & $M^{(4,\,22)}_{k}$& $M^{(4,\,33)}_{k}$& $M^{(4,\,23)}_{k}$ \\ \hline
0   &0& 10.03& -2.51&   0.31 &0.049&   0.16\\ \hline 1
&0& 0   &0& 0 &0& 0\\ \hline 2   &36.32&   46.85 &-6.50
&-0.29&-0.015& -0.14\\ \hline 3&  0   &0& 0&  0&  0&  0\\ \hline 4&
-71.50    &-100.66& 16.09&    0.59 &0.029&
0.25\\ \hline 5   &-10.36   &-15.39 &2.66&    0.077
&0.0034&   0.032\\ \hline 6&  116.21 &170.73& -29.24&
-0.98&   -0.054&   -0.39\\ \hline 7&  45.69 &73.09&
-13.997 &-0.275&   -0.012&   -0.105\\ \hline 8&  -169.5&
-260.38& 47.12 &1.01    &0.055&   0.37\\ \hline 9
&-117.70  &-196.09 &39.54    &0.34    &0.016&
0.12\\ \hline 10& 219.56 &354.27& -67.36&   9.42E-07
&-9.59E-05&  2.20E-06\\ \hline 11& 234.61 &406.45& -85.03&
0.94&    0.042 &0.30\\ \hline 12& -240.62&   -406.38&
79.92&    -2.49&   -0.11&   -0.75\\ \hline 13& -398.72&
-724.58& 157.86&    -6.21 &-0.248&  -1.76\\ \hline 14&
198.27&    345.6 &-67.8&  5.33 &0.22&   1.42\\
\hline 15& 597.98 &1147.26& -261.54&   19.24&    0.66
&4.81\\ \hline 16& -57.49&   -83.78 &7.19    &-3.89&
-0.149 &-0.91\\ \hline 17& -797.17 &-1618.27& 386.31
&-43.42&  -1.25&   -9.41\\ \hline 18& -208.15&   -471.79
&132.09&-11.56 &-0.26   &-2.31\\ \hline 19& 935.74&
2010.21& -501.69&   77.96 &1.81&   14.30\\ \hline 20&
604.56&    1387.68& -379.52&   55.88 &1.1&   9.31\\
\hline
\end{tabular}
\end{ruledtabular}
\end{table}

\begin{table}[h]
\caption{\label{Table3} Numerical coefficients for $\mu_6$ for a gas of viscoelastic particles [Eqs.~(
\ref{mupw})].}
\begin{ruledtabular}
\begin{tabular}{ | l | l | l | l | l | l | l |}
     \hline
          $k$ & $M^{(6,\,0)}_{k}$ & $M^{(6,\,2)}_{k}$ & $M^{(6,\,3)}_{k}$ & $M^{(6,\,22)}_{k}$& $M^{(6,\,33)}_{k}$& $M^{(6,\,23)}_{k}$ \\ \hline
0&  0&  112.80&    -84.60&    -4.93&   0.022&    -0.82\\
\hline 1&  0&  0& 0   &0& 0   &0\\ \hline 2&  209.94&    633.78&
-245.02&   16.75 &0.26& 2.23\\ \hline 3&  0   &0  &0& 0   &0
&0\\ \hline 4&  -525.04&    -1718.36& 717.54&    -48.53&-0.45&
-4.85\\ \hline 5&  -61.21 &-203.84& 86.30 &-7.2
&-0.061   &-0.60\\ \hline 6   &1097.057&   3731.6 &-1607.5&
133.88 &0.64    &8.93\\ \hline 7&  338.22 &1206.7
&-544.06   &46.04    &0.18    &2.3\\ \hline 8&  -2042.4&
-7140.7& 3126.2  &-347.5&   -0.75    &-11.58\\ \hline 9
&-1116.58   &-4131.0 &1913.43   &-196.42   &-0.21
&-3.27\\ \hline 10& 3412. &12219.7 &-5395.76&  806.66&
-0.0022&   -0.0012\\ \hline 11& 2856.1 &10832.9 &-5076.7
&683.4    &-0.647&  -11.39\\ \hline 12& -5055.2    &-18485.3
&8165.6 &-1634.9&   3.29 &54.5\\ \hline 13& -6187.9
&-24012.3 &11320    &-2006.6   &5.40 &100.33\\ \hline 14&
6435.9 &23886 &-10417.   &2818.3&   -10.83
&-187.9\\ \hline 15& 11807 &46936.9&-22224.5   &5062.8
&-21.37   &-421.9\\ \hline 16& -6433.2    &-23805. &9879.&
-3867.&   21.14& 386.7\\ \hline 17& -20203.4    &-82418.7
&39185.8    &-11160.6   &60.27 &1302.07\\ \hline 18& 3203.3
&10084 &-2516.6&  3251.2 &-23.5&  -433.49\\ \hline 19&
31188.6&    130743. &-62400.8    &21721.9    &-133.14
&-3258.3\\ \hline 20& 5784.3& 29433.95& -17812.7&   2307.9
&-6.45&  -384.65\\ \hline
\end{tabular}
\end{ruledtabular}
\end{table}

\newpage

\begin{table}[h]
\caption{\label{Table4} Numerical coefficients for $\mu_4$ and $\mu_6$ for a gas with a constant restitution
coefficient [Eqs.~ (\ref{muconst})].}
\begin{ruledtabular}
\begin{tabular}{ | l | l | l | l | l | l | l |}
     \hline
$k$ & $N^{(4,\,0)}_{k}$ & $N^{(4,\,2)}_{k}$ & $N^{(4,\,3)}_{k}$ & $N^{(4,\,22)}_{k}$& $N^{(4,\,33)}_{k}$& $N^{(4,\,23)}_{k}$ \\ \hline
0&147456&277504&-46336&2192&455&1096\\ \hline
1&-147456&-211968&29952&-144&-135&-72\\ \hline
2&32768&30720&-2560&-480&-50&-240\\ \hline
3&-32768&-30720&2560&480&50&240\\ \hline
\end{tabular}
\end{ruledtabular}
\begin{ruledtabular}
\begin{tabular}{ | l | l | l | l | l | l | l |}
     \hline
$k$ & $N^{(6,\,0)}_{k}$ & $N^{(6,\,2)}_{k}$ & $N^{(6,\,3)}_{k}$ & $N^{(6,\,22)}_{k}$& $N^{(6,\,33)}_{k}$& $N^{(6,\,23)}_{k}$ \\ \hline
0&-1884160&-8496128&4311296&26448&-3113&4408\\ \hline
1&1884160&7054336&-3229952&10416&1193&1736\\ \hline
2&-720896&-2920448&1401856&52032&2396&8672\\ \hline
3&720896&2396160&-1008640&-2880&-860&-480\\ \hline
4&-131072&-286720&71680&-13440&-280&-2240\\ \hline
5&131072&286720&-71680&13440&280&2240\\ \hline
\end{tabular}
\end{ruledtabular}
\end{table}


\newpage

\begin{thebibliography}{99}

\bibitem{Herr1998} {\it Physics of Dry Granular Media},  edited by H. J. Herrmann, J.-P. Hovi, and
S. Luding, NATO ASI Series (Kluwer, Dordrecht, 1998).

\bibitem{jnb96} H.M. Jaeger, S.R. Nagel, and R.P. Behringer, Rev. Mod. Phys. \textbf{68}, 1259 (1996).

\bibitem{hw04} {\it The Physics of Granular Media}, edited by  H. Hinrichsen,
and D.E. Wolf, (Wiley, Berlin, 2004).

\bibitem{r00} G.H. Ristow, {\it Pattern Formation in Granular Materials} (Springer-Verlag, Berlin, 2000).

\bibitem{d00} J. Duran, {\it Sands, Powders and Grains} (Springer-Verlag, Berlin, 2000).

\bibitem{PL2001} {\it Granular Gases}, edited by T. Poeschel and S. Luding, Lecture Notes in Physics Vol. 564
(Springer, Berlin, 2001).

\bibitem{PB2003} {\it Granular Gas Dynamics}, edited by T. Poeschel and N. V. Brilliantov, Lecture Notes
in Physics Vol. 624 (Springer, Berlin, 2003).

\bibitem{bp04} N.V. Brilliantov, and T. P\"oschel, {\it Kinetic Theory of Granular Gases} (Oxford University Press, Oxford, 2004).

\bibitem{g03} I. Goldhirsch, Ann. Rev. Fluid Mech. \textbf{35}, 267 (2003).

\bibitem{Brahic1984} {\it Planetary Rings}, edited by R. Greenberg and A. Brahic
(Arizona University Press, Tucson, 1984).

\bibitem{h83} P.K. Haff, J. Fluid Mech. \textbf{134}, 401 (1983).

\bibitem{gz93} I. Goldhirsch, and G. Zanetti, Phys. Rev. Lett. \textbf{70}, 1619 (1993).

\bibitem{dp03} S.K. Das, and S. Puri, Europhys. Lett. \textbf{61}, 749 (2003).

\bibitem{dppre03} S.K. Das, and S. Puri, Phys. Rev. E \textbf{68}, 011302 (2003).

\bibitem{n03} H. Nakanishi, Phys. Rev. E \textbf{67}, 010301 (2003).

\bibitem{nebo79} T.P.C. van Noije,  M.H. Ernst,  R. Brito, and J.A.G. Orza, Phys. Rev. Lett. \textbf{79}, 411 (1997).

\bibitem{neb98} T.P.C. van Noije, M.H. Ernst, and R. Brito, Phys. Rev. E \textbf{57}, R4891 (1998).

\bibitem{ap06} S.R. Ahmad, and S. Puri, Europhys. Lett. \textbf{75}, 56 (2006).

\bibitem{ap07} S.R. Ahmad, and S. Puri, Phys. Rev. E \textbf{75}, 031302 (2007).

\bibitem{Sant2000} J. M. Montanero and A. Santos, Granular Matter,  \textbf{2}, 53 (2000).

\bibitem{vne98} T.P.C. van Noije and M.H. Ernst, Granular Matter \textbf{1}, 57 (1998).

\bibitem{shear} L. Bocquet, W. Losert, D. Schalk, T. C. Lubensky, and J. P. Gollub, Phys. Rev. E \textbf{65}, 011307 (2001).

\bibitem{wildman} R. D. Wildman, and D. J. Parker, Phys. Rev. Lett \textbf{88}, 064301 (2002).

\bibitem{menon} K. Feitosa, and N. Menon, Phys. Rev. Lett \textbf{88}, 198301 (2002).

\bibitem{rotdriv} O. Zik, D. Levine, S. G. Lipson, S. Shtrikman, and J. Stavans, Phys. Rev. Lett \textbf{73}, 644 (1994).

\bibitem{puha} S. Puri and H. Hayakawa, Physica A \textbf{270}, 115 (1999); \textbf{290}, 218 (2001).

\bibitem{electrodriven} I. S. Aranson, and J. S. Olafsen, Phys. Rev. E \textbf{66}, 061302 (2002).

\bibitem{magndriven} A. Snezhko, I. S. Aranson, and W.-K. Kwok, Phys. Rev. Lett.
\textbf{94}, 108002 (2005).


\bibitem{spahn1997} F. Spahn, U. Schwarz, and J. Kurths,  Phys. Rev. Lett. \textbf{78}, 1596 (1997).

\bibitem{ringbook} J. Schmidt, K. Ohtsuki, N.Rappaport, H. Salo, and F. Spahn. \textit{Dynamics of Saturn's
dense rings}. In: M.K. Dougherty, L.W. Esposito, and S.M. Krimigis (Eds.),
\textit{Saturn from Cassini –- Huygens}. Springer, pp. 413–458 (2009)

\bibitem{wm} D.R.M. Williams,  and F.C. MacKintosh, Phys. Rev. E \textbf{54}, R9 (1996);
D.R.M. Williams, Physica A \textbf{233}, 718 (1996).

\bibitem{w60} W. Goldsmit, {\it The Theory and Physical Behavior of Colliding Solids} (Arnold, London, 1960).

\bibitem{bhd84} F.C. Bridges, A. Hatzes, and D.N.C. Lin, Nature \textbf{309}, 333 (1984).

\bibitem{kk87} G. Kuwabara, and K. Kono, J. Appl. Phys. Part 1 \textbf{26}, 1230 (1987).

\bibitem{titt91} T. Tanaka, T. Ishida, and Y. Tsuji, Trans. Jap. Soc. Mech. Eng. \textbf{57}, 456 (1991).

\bibitem{rpbs99} R. Ramirez, T. P\"oschel, N.V. Brilliantov, and T. Schwager, Phys. Rev. E \textbf{60}, 4465 (1999).

\bibitem{bshp96} N.V. Brilliantov, F. Spahn, J.M. Hertzsch, and T. P\"oschel, Phys. Rev. E \textbf{53}, 5382 (1996).

\bibitem{mo97} W.A.M. Morgado, and I. Oppenheim, Phys. Rev. E \textbf{55}, 1940 (1997).

\bibitem{sp98} T. Schwager, and T. P\"oschel, Phys. Rev. E \textbf{57}, 650 (1998).

\bibitem{sp08} T. Schwager, and T. P\"oschel, Phys. Rev. E \textbf{78}, 051304 (2008).

\bibitem{TrizackKrapPRL} E. Trizac, and P. L. Krapivsky Phys. Rev. Lett., \textbf{91}, 218302 (2003)

\bibitem{adh2007} N.V. Brilliantov, N. Albers, F. Spahn, and T. P\"oschel, Phys. Rev. E \textbf{76}, 051302 (2007).

\bibitem{adh2006} N.V. Brilliantov, and F. Spahn, Math. Comp. Simul. \textbf{72}, 93 (2006).

\bibitem{gs95} A. Goldshtein, and M. Shapiro, J. Fliud Mech. \textbf{282}, 75 (1995).

\bibitem{ep97} S.E. Esipov, and T. P\"oschel, J. Stat. Phys. \textbf{86}, 1385 (1997).

\bibitem{bp06} N.V. Brilliantov, and T. P\"oschel, Europhys. Lett. \textbf{74}, 424 (2006).

\bibitem{Sant2009} A. Santos and J. M. Montanero, Granular Matter,  \textbf{11}, 157 (2009).

\bibitem{hob00} M. Huthmann, J.A.G. Orza, and R. Brito, Granular Matter \textbf{2}, 189 (2000).

\bibitem{bcr96} J.J. Brey, M.J. Ruiz-Montero,  and D. Cubero, Phys.  Rev. E \textbf{54}, 3664 (1996).

\bibitem{BrilPRL2004} N. Brilliantov, C. Saluena, T. Schwager, and T. Poschel,
Phys. Rev. Lett. \textbf{93}, 134301 (2004).

\bibitem{bp00} N.V. Brilliantov, and T. P\"oschel, Phys. Rev. E \textbf{61}, 5573 (2000).

\bibitem{bb09} A.S. Bodrova, and N.V. Brilliantov, Physica A \textbf{388}, 3315 (2009).

\bibitem{Compbook} T. P\"oschel, and T. Schwager, {\it Computational Granular Dynamics} (Springer, Berlin, 2005).


\bibitem{brey2004} J.J. Brey, and M.J. Ruiz-Montero,  Phys. Rev. E \textbf{69}, 011305 (2004).


\bibitem{Pag1999} T. P. C. van Noije, M. H. Ernst, E. Trizac and I. Pagonabarraga, Phys. Rev. E \textbf{59}, 4326 (1999).

\end{thebibliography}
\end{document}